\documentclass{article}
\usepackage[utf8]{inputenc} 
\usepackage{url}            
\usepackage{amsfonts}       
\usepackage{graphicx}
\usepackage{tabularx}
\usepackage{amsmath,amssymb}
\usepackage{amsthm}
\usepackage{algorithm}
\usepackage{algorithmicx}
\usepackage{algpseudocode}
\newtheorem{definition}{Definition}
\newcommand{\spara}[1]{\smallskip\noindent{\bf{#1}}}

\begin{document}
\title{Community Search in Attributed Networks using Dominance Relationships and Random Walks}

\author{
    Nikolaos Georgiadis \\
    Aristotle University of Thessaloniki \\
    \texttt{ngeorgii@csd.auth.gr}
    \and
    Eleftherios Tiakas \\
    International Hellenic University \\
    \texttt{tiakas@ihu.gr}
    \and
    Apostolos N. Papadopoulos \\
    Aristotle University of Thessaloniki \\
    \texttt{papadopo@csd.auth.gr}
}

\date{}
\maketitle

\begin{abstract}
    Community search in attributed networks poses a dual challenge: balancing \textit{structural connectivity}---the network's topological properties---and \textit{attribute similarity}---the shared characteristics of nodes. This paper introduces a novel algorithm that integrates hop-based and random-walk-based methods to identify high-quality communities, effectively addressing this balance. Our approach employs the concept of the domination score to quantify the influence of nodes based on their attributes, followed by $k$-core extraction to ensure strong structural cohesion within the communities. By considering both the network structure and node attributes, the algorithm identifies communities that are not only well-connected, but also share meaningful attribute similarities. We evaluated the algorithm on large real-world datasets, demonstrating its ability to efficiently identify cohesive communities, making it suitable for applications such as social network analysis and recommendation systems.
\end{abstract}

%
%
\section{Introduction}\label{sec:intro}
The study of real-life networks leads to important research directions with many interesting application domains~\cite{barabasi2016}. Real-life networks tend to form communities, which are defined briefly as coherent subgraphs with an increased number of intra-subgraph links and a small number of inter-subgraph links.

A network is represented as a graph $G(V,E)$ where $V$ is the set of nodes or vertices and $E$ is the set of edges or links. In many cases, nodes contain attributes of different types. These attributes may play an important role in several downstream data mining tasks, such as classification, link prediction, network reconstruction, and community detection.

Community search in attributed networks is an important problem with many real-world applications.
In most application-specific networks, node attributes are used to record application data, which may affect their structure in future changes. Especially in dynamic networks, the node attributes play an important role for their evolution.

There are numerous fields where the targeted community search leads to understanding the structure and dynamics of these networks, which is critical, like marketing, e-commerce, transportation, social networks, citations, public health, cyber-security, and many others. Often, those real-world networks are rich in heterogeneous node-attribute information, such as user profiles in social networks, functional annotations in biological networks, or metadata in citation graphs.

However, traditional methods often focus solely on global structures or topological information, overlooking a crucial aspect, the existing node attributes. Local community detection based on both topological information and node attributes can provide concrete applications in the aforementioned fields, like: (i) targeted recommendations in e-commerce or content platforms, (ii) localized interventions in public health, (iii) Fine-grained security analysis in communication networks, and many others.

The problem we tackle in this work is briefly defined as follows. Let $G(V,E)$ be an undirected graph with a set of nodes $V$ with node attributes and a set of edges $E$.
Let also $v_0$ be a specific user-selected node, defined as the \textit{query node} and $f(S)$ be a goodness measure (score) of a subgraph $S$ that depends both on the cohesiveness of the subgraph and the values of node attributes. We need to detect a subgraph $S$ of $G$ such that:

\begin{itemize}
    \item $v_0 \in S$
    \item the score is $f(S)$ optimized.
\end{itemize}

Note that the problem can be generalized towards taking into account more query nodes simultaneously. However, in this work, we tackle the simplest version of the problem which includes just one query node.

\begin{figure}[hbt!]
    \centering
    \includegraphics[width=0.9\textwidth]{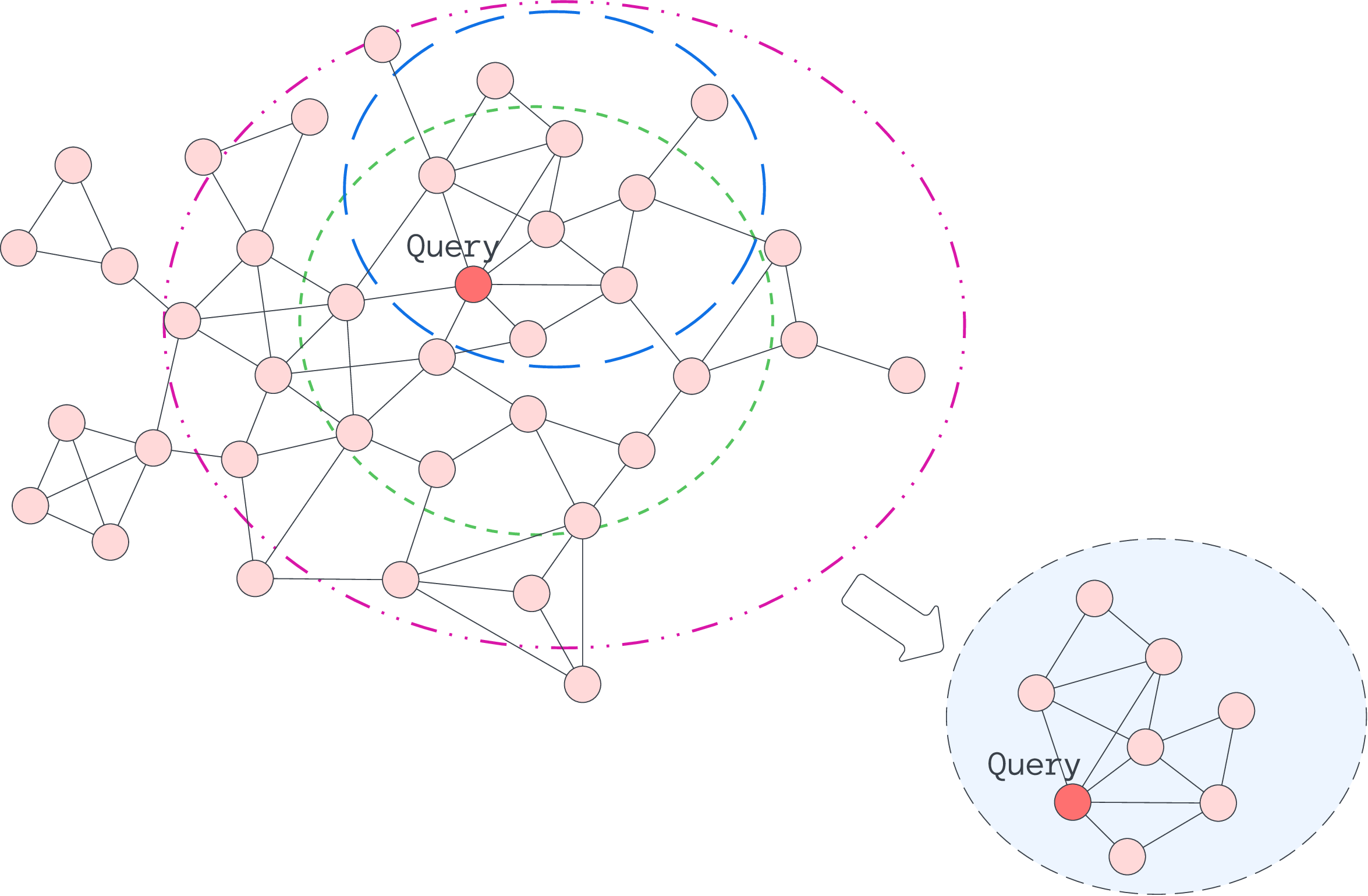}
    \caption{Example of community search: the query node is highlighted in red. The different subgraphs that may be defined are shown in the dashed circular areas. The selected subgraph is shown in the bottom right corner.}\label{fig:base_example}
\end{figure}

An example is shown in Figure~\ref{fig:base_example}. There are several strategies that we can follow to select the best subgraph $S$ that contains the query node. The selected subgraph shown in the figure is the one that maximizes the score $f(S)$ which is based on both the structural properties of the graph and the attribute values of the nodes.

The rest of the work is organized as follows. Related work in the area is briefly described in Section~\ref{sec:related}. The proposed methodology is given in detail in Section~\ref{sec:proposed}. Performance evaluation results are given in Section~\ref{sec:performance}. Finally, Section~\ref{sec:conclusions} summarizes our work and highlights some interesting future research directions.
%
%

%
%
\section{Related Work}\label{sec:related}

The problem of local communities detection, i.e.\ communities in a graph/network to which a given starting node belongs, has gained a lot of attention in the last decades, and several methods and algorithms have been proposed.

Clauset~\cite{clauset2005local} proposed a measure of local community structure, which called local modularity, and an agglomerative algorithm that maximizes the local modularity in a greedy fashion for adding nodes to the resulted community.

Luo et al.~\cite{luo2008exploring} proposed three algorithms (greedy, KL-like, add-all) for finding local optimal community structures starting from a given node. A quality function that takes into account the edges within a community and the edges between communities is used.

Chen et al.~\cite{5231879} introduced a measure of local community structure that takes the connection among nodes in the community and the connection between communities into consideration. The proposed algorithm has two-phases that extracts all possible candidates first, and then optimizes the community hierarchy. To solve the outliers problem, the algorithm checks for the changes in a quality function of the community after removing nodes.

Wu et al.~\cite{wu2012link} proposed a three-phase algorithm that analyses link similarity for candidate nodes. The first phase adds the nodes with the largest similarity to the community, in a greedy fashion. The second phase checks the boundary nodes for remaining in the community, and the third phase removes the nodes that have more neighbors in the boundary than in the community.

Zhang and Wu~\cite{6425598} proposed a method to detect the local community of a given node, which first it finds the core node of the community and then expand the core node's cliques to construct the community.

Fanrong et al.~\cite{Fanrong2014} proposed a local community detection algorithm based on maximum clique extension (LCD-MC). The algorithm detects the set of all maximum cliques containing the source node, which are initialized as the starting local communities. Then, the maximum cliques are assigned into a community by greedy optimization until a certain objective is satisfied.

Interdonato et al.~\cite{interdonato2017multilayer} introduced the problem of local community detection for multilayer networks. The proposed Multilayer Local Community Detection framework (ML-LCD) conducts optimization with associated objective functions, which correspond to different ways to incorporate within-layer and across-layer topological features.

Luo et al.~\cite{8306896} proposed two local community detection algorithms, which are based on dynamic membership functions in order to detect and add the nodes with the greatest neighborhood intersect rate that are closed to the community, and some nodes that should not be omitted.

Luo et al.~\cite{8818674} proposed a multi-scale local community detection algorithm, which is based on local and global modularities.

Lyu et al.~\cite{8809875} proposed the evolutionary based local community detection algorithm (ELCD), which performs modularity optimization where the classic modularity concept is modified to adjust its application from the global community detection to the local community detection.

Luo et al.~\cite{LUO2020377} proposed a local community detection algorithm (LCDNN), which is based on NGC nodes (nearest nodes with greater centrality) and adopts the fuzzy relation in local community detection that measures the closeness from nodes to their corresponding NGC nodes.

Guo et al.~\cite{guo2022local} proposed a local community detection algorithm based on local modularity density, which has a core area detection phase and a local community extension phase. In the core area detection phase the modularity density is used as a measure of quality. In the local community extension phase, both the influence and the similarity of the nodes are considered, in order to determine boundary nodes and to reduce the sensitivity to seed node selection.

Wu et al.~\cite{Wu2015} studied the ``free rider effect'', which is the appearance of irrelevant subgraphs in local community detection. They proposed a query biased node weighting scheme to reduce the free rider effect, a density metric to integrate the edge and node weights, and efficient algorithms for local community detection based on the query biased densest connected subgraph.

Zakrzewska and Bader~\cite{7403595} proposed a local community detection algorithm in dynamic graphs where new nodes and edges may be inserted or old may be removed. The algorithm is based on dynamic seed set expansion, which incrementally updates the community as the underlying graph changes.

Yao et al.~\cite{yao2017community} proposed a variable influence local community detection algorithm (VI), which analyzes the influence of nodes and constructs and resizes the community involving nodes with high influence according to users' demands. Yao et al.~\cite{yao2017community} proposed
also a similar algorithm for the case of starting from a specific given node.

Ding et al.~\cite{DING2018188} proposed a method called Robust Two-stage Local Community Detection (RTLCD), which in the first stage of seed selection searches the core member of the community, and in the second stage of community extension it expands from the core member by taking into account the node mass and the node relation strength.

Hollocou et al.~\cite{Hollocou2018} proposed the MULTICOM algorithm, which detects multiple communities nearby a given seed set S. The main strategy of the algorithm is to define an appropriate graph embedding around the seed set using local scoring metrics. Then new seeds are picked to recover multiple communities.

Zhang et al.~\cite{8727765} proposed a Local Community Detection method based on network Motif (LCD-Motif), which conducts a local expansion of a seed set to identify the local community with minimal motif conductance, by using a generalization of the conductance metric for network motifs.

Luo et al.~\cite{LuoBian2020} proposed a random walk model, RWM, to detect local communities in multiple networks for a given query node set. The algorithm can detect all query-relevant local communities for multi-domain networks. In this work two more approximation methods were proposed that improve the computational efficiency.

Wang et al.~\cite{wang2021local} proposed a local community detection method based online graph through degree centrality and expansion (LCDDCE). First, the proposed algorithm employs a model and transfers the edges of the original graph into nodes of a new graph (line graph). Then, the nodes are ranked by a similarity score and seeds are derived. Finally, the community is constructed and expanded by using a fitness function.
~\\

\spara{Main differences with the proposed approach}

All previous works are applied on simple structured networks and graphs, static or dynamic, small or large, except the work of Interdonato et al.~\cite{interdonato2017multilayer}, which is applied in multilayer networks, and the work of Luo et al.~\cite{LuoBian2020} which is applied in multiple networks. In our work, we consider attributed graphs, that is, graphs that have a specific number of attributes in the nodes, and we also consider the domination relationships between the nodes according to these attributes and the maximum $k$-core for the resulted communities.

To the best of our knowledge, this is the first work that applies domination computations to the local community search problem, expanding the applications in the community detection field. Moreover, we also propose an alternative that utilizes random walks, in order to further improve the performance using also a different way to describe the content of a node.
%
%

%
%
\section{Proposed Methodology}\label{sec:proposed}
All the basic symbols used in this and the following sections are illustrated in table~\ref{tab:symbols}.
\begin{table}[ht]
    \centering
    \begin{tabular}{c m{8cm}}
        \hline
        \textbf{Symbol} & \textbf{Description}                                                              \\ \hline
        \( G \)         & The Graph                                                                         \\ \hline
        \( n_x \)       & Current derived node                                                              \\ \hline
        \( d \)         & The maximum node degree                                                           \\ \hline
        \( s \)         & The grid size parameter                                                           \\ \hline
        \( \gamma \)    & The maximum number of objects that one cell holds                                 \\ \hline
        \( h \)         & The number of hops in the hop-search algorithm                                    \\ \hline
        \( m \)         & Total number of attributes/dimensions in nodes                                    \\ \hline
        \( p \)         & The length of the path parameter for the random walk-search algorithm             \\ \hline
        \( w \)         & The number of iterations parameter for the random walk-search algorithm           \\ \hline
        \( \sigma \)    & The evaluation metric used to determine the rank of the community                 \\ \hline
        \( dom_i \)     & The domination score of the node in position \( i \)                              \\ \hline
        \( MAX_{dom} \) & The maximum domination score of the subgraph in which the evaluation is performed \\ \hline
        \( N \)         & The total number of nodes of the evaluated community                              \\ \hline
    \end{tabular}
    \caption{Table of Symbols}\label{tab:symbols}
\end{table}

\subsection{Background}\label{sec:background}

\subsection*{Attributed Graphs}
An attributed graph is a specialized type of graph where each node has a set of attributes~\cite{Yang2013}. These attributes provide additional information about the graph that complements the connections between nodes. For example in a social network, nodes representing individuals might have attributes such as age, location, interests or occupation. Similarly in a citation network, nodes representing research papers could have attributes like publication count, h-index, keywords or author affiliations. The key characteristic is that each node $v$ in the graph $G = (V, E)$ has a vector of attribute values $X_v = (x_{v1}, x_{v2}, \ldots, x_{vd})$, where $d$ represents the total number of distinct attributes.

Including these attributes we can perform a more detailed analysis than traditional methods that only look at connections. While standard network analysis is good for understanding the structure of a network, it misses the specific characteristics of the nodes. Using attributed graphs allows us to find groups and patterns that are not only well-connected but also share meaningful traits. This is essential for solving complex problems where both the network structure and the features of the nodes are important.

\subsection*{Domination Relationships and Score}
A central idea in this research is that some nodes in a network can be considered more important or ``dominant'' than others based on their specific features. We determine this dominance by calculating a ``domination score''. This score is a number that shows how much better one node's features are compared to others, allowing us to rank them~\cite{OurBook}.

The domination score offers a way to find local communities within networks. Instead of just looking for groups of nodes that are similar or well-connected, it is also considering the  dominant nodes that are participating in the community. This helps us find communities that form around key, influential members, providing a more detailed picture of how these groups are structured.

Without loss of generality, assuming that in each node feature greater values are preferred, we define the dominance and domination score as follows:

\begin{definition}[dominance]\label{ch1.def.dominance}
    An object $p=(p.x_1,p.x_2,\ldots,p.x_d) \in \mathbb{R}^d$ dominates another object $q=(q.x_1,q.x_2,\ldots,q.x_d) \in \mathbb{R}^d$,
    i.e., $p \succ q$, when: $\forall i \in \{1,\ldots,d\} : p.x_i \ge q.x_i \wedge \exists i \in \{1,\ldots,d\} : p.x_i > q.x_i$.
    This means that $p$ is as good as $q$ in all dimensions, and it is strictly better than $q$ in at least one dimension.
\end{definition}

\begin{definition}[domination score]\label{ch1.def.domscore}
    The domination score of a point $p$, $dom(p)$ is defined as: $dom(p)=|\{q \in D : p \succ q\}|$. A top-$k$ dominating query returns the $k$ objects with the highest domination scores.
\end{definition}

\subsection*{Extraction of $k$-core}
The concept of $k$-core is a fundamental notion in graph analysis which is used to identify structurally cohesive subgraphs within a network~\cite{malliaros2020}. The $k$-core of a graph is defined as a maximal connected subgraph in which all nodes have a degree of at least $k$. In other words, to be part of a $k$-core, a node must be connected to at least $k$ other nodes \textit{within the same subgraph}. The process of finding $k$-cores, often referred to as $k$-core decomposition, involves the iterative removal of all nodes with a degree less than $k$ until no such nodes remain. This results in a nested hierarchy of cores, where the $(k+1)$-core is always a subgraph of the $k$-core, representing increasingly dense and tightly connected regions of the network.

In the context of community search, and particularly for the algorithms discussed in this paper, $k$-core extraction serves as a powerful mechanism to ensure strong structural cohesion within identified communities. By applying $k$-core decomposition to a candidate subgraph or the entire graph, the algorithms can isolate and retain only those nodes that satisfy a minimum connectivity threshold. Extracting the maximum $k$-core (i.e., the core corresponding to the highest possible value of $k$ for which a non-empty subgraph exists) of a given community or subgraph is particularly useful for identifying its most densely connected and robust portion. This helps filter out peripheral or loosely connected nodes, thereby refining the community structure to highlight its most integral members and ensuring that the discovered communities possess strong underlying structural integrity.

\subsection{Overview}\label{sec:algorithm}
In this section we present an overview of the two algorithms used for the community search: the hop-search and the random walk-search algorithm. Both methods try to find densely connected subgraphs, each using a different approach.

The hop-search algorithm focuses on constructing a local view of the network around a query node. This is achieved by creating a subgraph that includes all nodes within a specified number of hops from the query node (egonet). Then within this egonet, the algorithm calculates the domination score for each node based on the node attributes and sorts them in descending order. Finally, the algorithm forms communities starting from the highest-scoring node and extracts firstly the egonet of the node and then the maximum k-core from this egonet.

The random walk-search algorithm uses random walks to sample the network. Starting from the query node, it performs multiple random walks to generate an induced subgraph. Then the algorithm calculates the domination score of each node based on the node attributes and sorts them in descending order. As a final step the algorithm forms communities starting from the highest-scoring node and for each node, it again extracts the induced subgraph of the random walk and then finds the maximum k-core of this subgraph.

Both algorithms aim to identify cohesive and well-defined communities by combining node attributes and structural properties of the graph.

\subsection{Hop-Based Algorithm (HBA)}
The Hop-Based Algorithm (HBA), (Algorithm~\ref{alg:hop_search_algorithm}), begins by extracting the induced subgraph (egonet) with a distance of $h$+1 from the query node. This step ensures that the subgraph includes all nodes within the specified number of hops from the query node, effectively creating a localized view of the network. Following this, the algorithm calculates a domination score for each node within this subgraph. The domination score derived from the attributes of the nodes serves as a measure of each node's importance within the subgraph. Once these scores are computed, the nodes are sorted in descending order according to their domination scores prioritizing the nodes with the highest domination score.

\begin{algorithm}[!b]
    \caption{Hop-Based Algorithm (HBA)}\label{alg:hop_search_algorithm}
    \begin{algorithmic}[1]
        \Require $G$ (graph), $Query$ (The query node), $h$ (\# of hops)
        \Ensure Results
        \State Get the egonet with distance $h + 1$ starting from the $Query$ node
        \State Get the attributes of the nodes from the induced egonet subgraph
        \State Calculate the domination score for each node
        \State Sort by score in descending order
        \For{each node $n_x$ starting from the one with the highest domination score}
        \State Get the egonet of the node $n_x$ with distance $h$
        \State Find the max k-core
        \State Get the induced subgraph of the max k-core
        \EndFor
        \State Calculate the stats of the subgraph
        \Return Results
    \end{algorithmic}
\end{algorithm}

Next the algorithm proceeds by focusing on the node with the highest domination score to form a community. For each node the induced subgraph of its egonet with distance hop is obtained. This is followed by the extraction of the induced subgraph of the maximum $k$-core, ensuring a densely connected community. This process results in a potential community for the node. Finally, a specific metric is calculated to evaluate the quality of the formed community. This metric helps in determining whether the identified community is substantial and cohesive.

\begin{figure}[!t]
    \centering
    \includegraphics[width=0.8\textwidth]{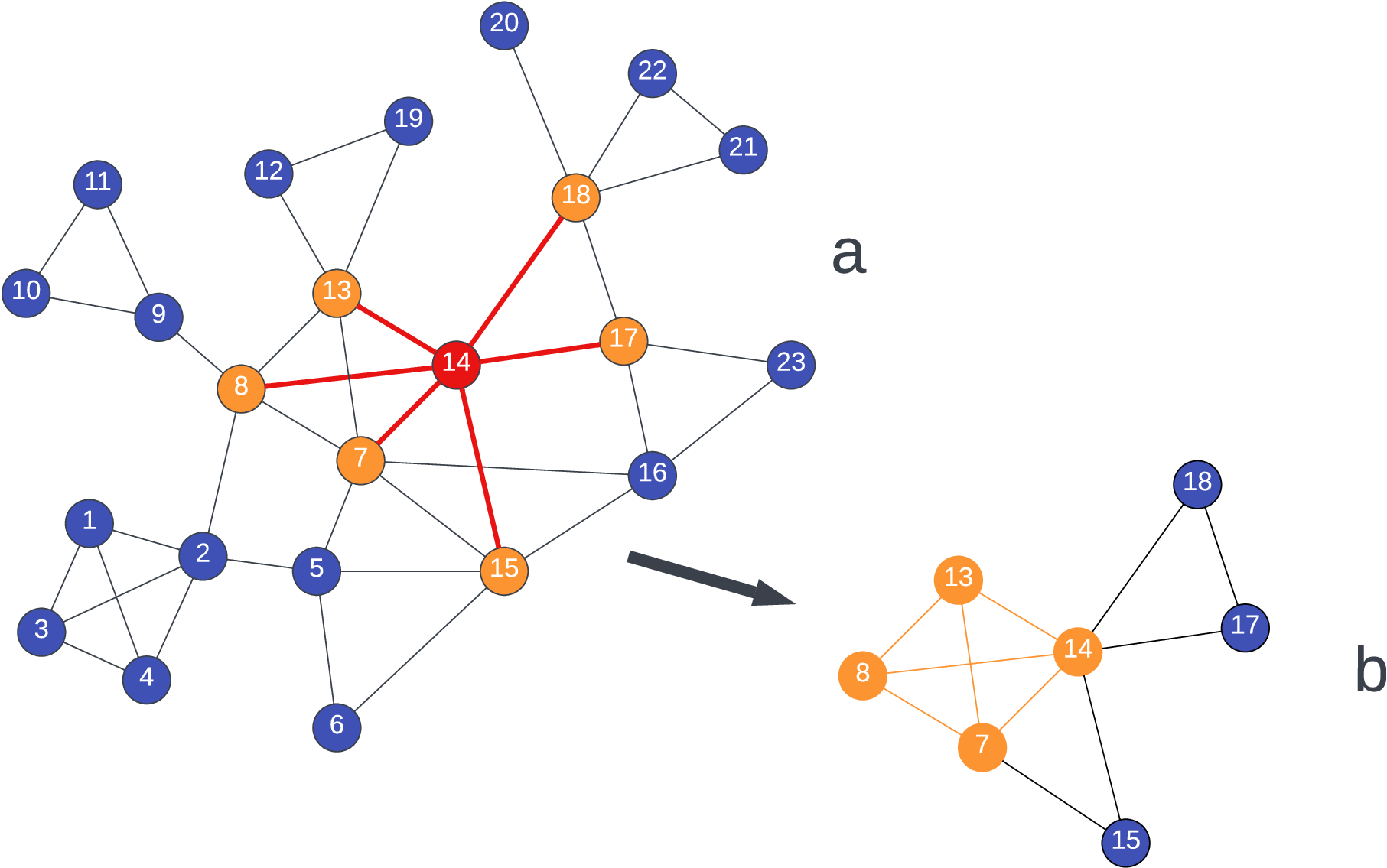}
    \caption{Example of HBA result}\label{fig:example-hop-search}
\end{figure}

Figure~\ref{fig:example-hop-search} illustrates the procedural steps for selecting and forming a community using the hop-search algorithm. The process begins with the initial node labeled 14. From this starting point the algorithm selects all nodes with distance of one hop from node 14 containing the immediate neighbors within the entire graph, as depicted in sub-figure (a). In sub-figure (b), the induced subgraph is constructed using the previously selected nodes. Within this induced subgraph, the next step involves identifying and selecting the nodes that are part of the highest k-core. In this case, the nodes that belong to the highest k-core of the induced subgraph are nodes 7, 8, 13, and 14.

\subsection{Random Walk-Based Algorithm (RWBA)}
The random walk-search algorithm, (Algorithm~\ref{alg:random_walk_search_algorithm}), operates using two key parameters: the random walk path length ($p$) and the total number of iterations ($w$). Initially, the algorithm generates an induced subgraph by executing the random walk with the specified path length and number of iterations. This process ensures that the subgraph represents a sample of the network, influenced by the random traversal of nodes and edges. Following the creation of this subgraph, the algorithm computes the domination score for each node based on their attributes. These scores quantify each node's prominence within the subgraph. After calculating the scores, the nodes are sorted in descending order prioritizing those with the highest domination scores.

The algorithm then forms communities by starting with the node that has the highest domination score. For each node, the induced subgraph generated from the random walk with the specified parameters is obtained again. This is followed by extracting the induced subgraph of the maximum k-core, ensuring a densely connected subset of the subgraph. The resulting subgraph is considered a potential community for the node. Finally, a specific metric is calculated to assess the quality of each formed community. This metric determines whether the community is well defined and cohesive, thus evaluating the effectiveness of the community formation process.

\begin{algorithm}[!t]
    \caption{Random Walk-Based Algorithm (RWBA)}\label{alg:random_walk_search_algorithm}
    \begin{algorithmic}[1]
        \Require $G$ (graph),
        $Query$ (the query node),
        $p$ (the length of the path for the random walk),
        $w$ (number of iterations for the random walk)

        \Ensure Results
        \State Get the subgraph $G'$ of the random walk search starting from the $Query$ node with parameters $p$ and $w$
        \State Get the attributes of the nodes from the induced subgraph
        \State Calculate the domination score for each node
        \State Sort by score in descending order
        \For{each node $n_x$ starting from the one with the highest domination score}
        \State Get the subgraph of the random walk search starting from the $n_x$ node with parameters $p$ and $w$ in $G'$
        \State Find the max k-core
        \State Get the induced subgraph of the max k-core
        \EndFor
        \State Calculate the stats of the subgraph
        \Return Results
    \end{algorithmic}
\end{algorithm}

\begin{figure}[!hbt]
    \centering
    \includegraphics[width=0.8\textwidth]{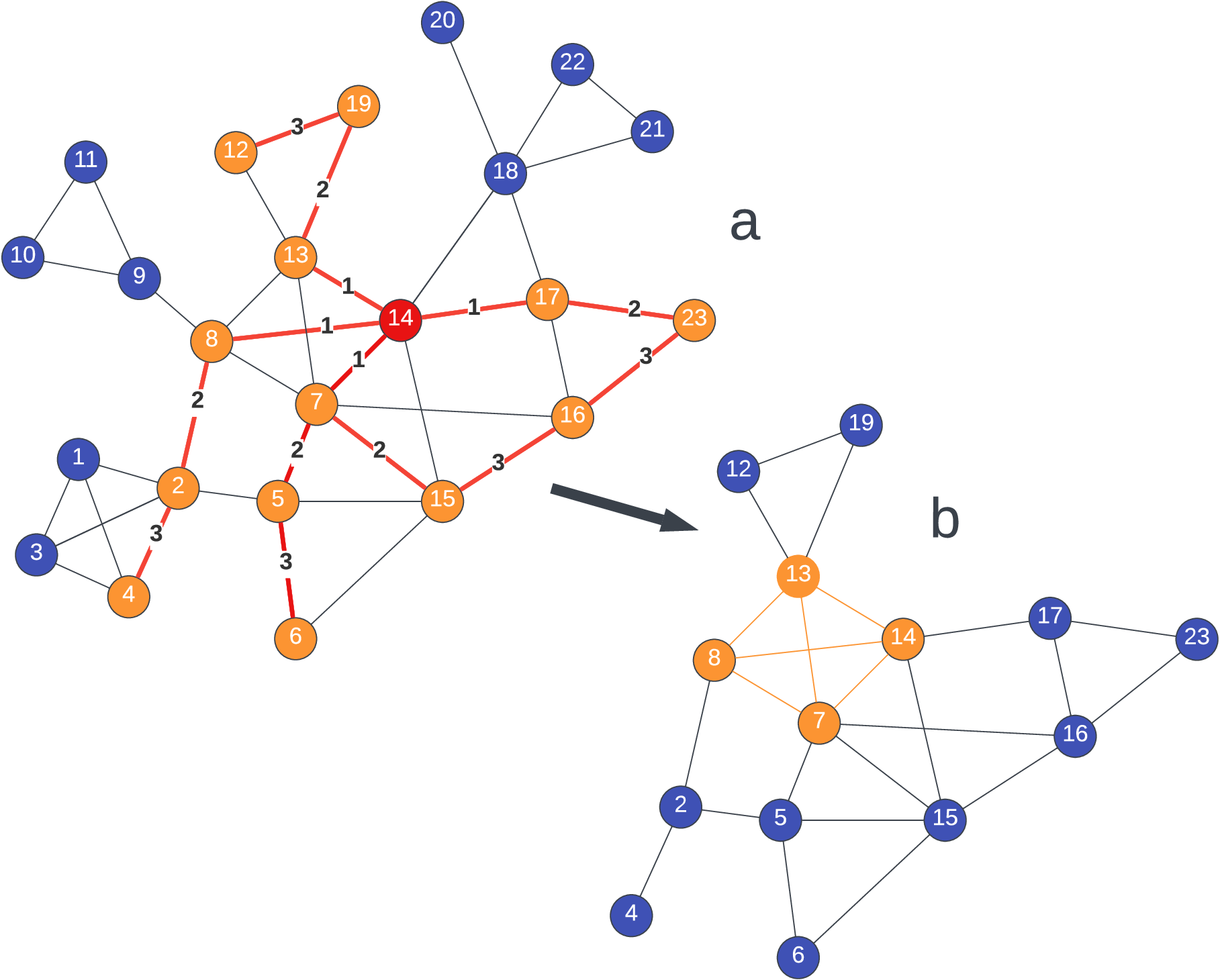}
    \caption{Example of random walk-search algorithm}\label{fig:example-random-walk-search}
\end{figure}

Figure~\ref{fig:example-random-walk-search}a illustrates the steps involved in the selection and formation of a community using the random walk-search algorithm. The process initiates from the initial node, labeled as node 14. From this starting point, a random walk of length 3 is performed, selecting all nodes included in the path. This procedure is repeated five times, resulting in an aggregation of nodes from multiple random walks.

The induced subgraph, shown in Figure~\ref{fig:example-random-walk-search}b, is formed using the nodes accumulated from these random walk paths. Within this induced subgraph, the next step involves identifying and selecting the nodes that belong to the highest k-core. In this scenario, the nodes that participate in the highest k-core of the induced subgraph are nodes 7, 8, 13, and 14. This approach ensures that the resulting community is not only a product of the random walk paths but also possesses a strong internal cohesion, as demonstrated by the k-core analysis.

\subsection{Theoretical Analysis}

\subsubsection{Complexity Analysis of HBA}
Let us denote as $d$ the maximum node degree of the initial graph $G$.
Let also $s$ be the grid size parameter, which defines in how
many cells a specific dimension is divided, $\gamma$ be the maximum
number of objects that one cell holds, and $m$ the total
number of attributes/dimensions in nodes of $G$.
Let also $h$ the number of hops in the algorithm.

In Line 1 we get the egonet $G'$ of the starting node with distance $h+1$, thus we have a worst case complexity of $O(1+d+d^2+...+d^h+d^{h+1})$ =
$O(\frac{d^{h+2}-1}{d-1})$ = $O(d^{h+2})$ (when all the derived nodes
are different and have the maximum graph degree).

Line 2 does not contribute to the complexity as the attributes of the
derived nodes are kept in main memory.

\begin{figure}[!hbt]
    \centering
    \includegraphics[width=0.5\textwidth]{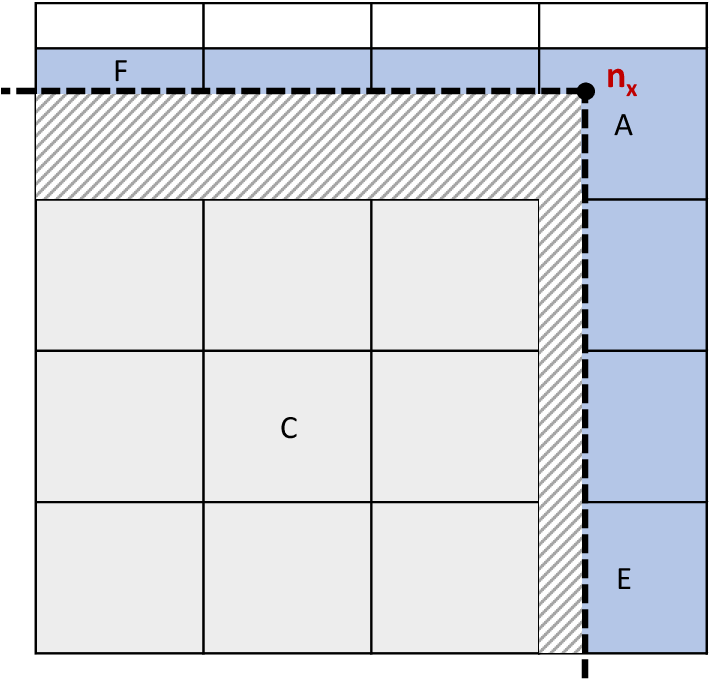}
    \caption{Example of the Grid that is used to calculate the domination score of node $n_x$. Cell $C$ is completely dominated while nodes that are included in cells $A$, $F$, and $E$ need to be evaluated}\label{fig:grid-example}
\end{figure}

In line 3 we calculate the domination score of the derived nodes of $G'$.
For each derived node $n_x$ we make a domination calculation as presented in Figure~\ref{fig:grid-example}.
The domination calculation directly receives the total number of objects from
the cells that are lying entirely inside the domination area of $n_x$
(this does not contribute to the complexity), and it conducts domination checks
for all objects that are lying in the corresponding cell lines ($m=2$) / planes ($m=3$) /
hyper-planes ($m>3$) that the projections of $n_x$ to the axis pass through.
In the worst case we may have to check a complete hyper-plane that contains
$s^{m-1}$ cells for each one of the $m$ dimensions, and we may have $\gamma$ objects
in each cell, thus we have a total number of $\gamma \times m \times s^{m-1}$
domination checks.

Therefore, for line 3 we have a total complexity of $O(\gamma \times m \times s^{m-1}
    \times d^{h+2})$.

The sort procedure in line 4 has a complexity of $O(d^{h+2} \log d^{h+2} )$ = \\
$O( (h+2)d^{h+2} \log d )$.

In the loop of line 5, for each derived node of $G'$ we get the egonet $G''$
with distance $h$, thus we have a worst case complexity of
$O(1+d+d^2+...+d^h)$ = $O(\frac{d^{h+1}-1}{d-1})$ = $O(d^{h+1})$,
Then we find the max k-core and get the induced graph, which in the
worst case can be a complete graph, thus the complexity is:
$O(\frac{d^{h+1}(d^{h+1}-1)}{2})$ = $O(d^{2h+2})$.
Therefore, for line 5 we have a total complexity: \\
$O(d^{h+2} \times (d^{h+1} + d^{2h+2}) )$.

The remaining calculations (lines 6 and 7) do not contribute more
to the total complexity.

Therefore, the total complexity of the algorithm is:

$O( d^{h+2} + \gamma m s^{m-1} \times d^{h+2} + (h+2)d^{h+2} \log d
    + d^{h+2} \times (d^{h+1} + d^{2h+2}) )$ =

$O( [ 1 + \gamma m s^{m-1} + (h+2)logd + d^{h+1} + d^{2h+2}  ] \times d^{h+2})$

As it holds that $(h+2)logd < d^{2h+2}$ and $d^{h+1} < d^{2h+2}$,
we derive the final  complexity: $O( (\gamma m s^{m-1} + d^{2h+2}  ) \times d^{h+2})$ =
$O( \gamma m s^{m-1} d^{h+2} + d^{3h+4})$.

\subsubsection{Complexity Analysis of the RWBA}
Let us denote by $d$ the maximum node degree of the initial graph $G$.
Let also $s$ be the grid size parameter, which defines in how
many cells a specific dimension is divided, $\gamma$ be the maximum
number of objects that one cell holds, and $m$ the total
number of attributes/dimensions in nodes of $G$.
Let also $p$ be the length of the path and $w$ the number of iterations for the random walk in the algorithm.

In line 1 we get the subgraph $G'$ of the random walk of the starting node with $p$-length paths and $w$ iterations, thus we have a worst case complexity of $O(p \times w)$ (when all the derived nodes from the paths are different).

Line 2 does not contribute to the complexity as the attributes of the
derived nodes are kept in main memory.

In line 3 we calculate the domination score of the derived nodes of $G'$.
For each derived node $n_x$ we make a domination calculation as presented in Figure~\ref{fig:grid-example}.
The domination calculation directly receives the total number of objects from
the cells that are lying entirely inside the domination area of $n_x$
(this does not contribute to the complexity), and conducts domination checks
for all objects that are lying in the corresponding cell lines ($m=2$) / planes ($m=3$) /
hyper-planes ($m>3$) that the projections of $n_x$ to the axis pass through.
In the worst case, we may have to check a complete hyper-plane that contains
$s^{m-1}$ cells for each one of the $m$ dimensions, and we may have $\gamma$ objects
in each cell, thus we have a total number of $\gamma \times m \times s^{m-1}$
domination checks.

Therefore, for line 3 we have a total complexity of $O(\gamma \times m \times s^{m-1} \times p \times w)$.

The sort procedure in line 4 has a complexity of $O(p \times w \times \log (p \times w) )$

In the loop of line 5, for each derived node of $G'$ we apply
a random walk with $p$ length paths and $w$ iterations, and we
get a subgraph $G''$, which in the worst case has $p \times w$ nodes.
In the derived subgraph $G''$ we find the max k-core and we get the
induced graph, which in the worst case can be a complete graph,
thus the complexity is:
$O(\frac{p \times w (p \times w - 1)}{2})$ = $O(\frac{p^2 \times w^2  - p \times w}{2})$
= $O(p^2 \times w^2)$.
Therefore, for line 5 we have a total complexity:
$O(p \times w \times p^2 \times w^2 )$ = $p^3 \times w^3$.

The remaining calculations (lines 6 and 7) do not contribute more
to the total complexity.

Therefore, the total complexity of the algorithm is:

$O( p \times w + \gamma \times m \times s^{m-1} \times p \times w + p \times w \times \log (p \times w) + p^3 \times w^3 )$ =

$O( [ 1 + \gamma \times m \times s^{m-1} + log(p \times w) + p^2 \times w^2  ] \times p \times w )$

As it holds that $log(p \times w) < p^2 \times w^2$,
we derive the final  complexity:

$O( (\gamma \times m \times s^{m-1} + p^2 \times w^2  ) \times p \times w )$ =
$O( \gamma \times m \times s^{m-1} \times p \times w + p^3 \times w^3)$.

%
%

%
%
\section{Performance Evaluation}\label{sec:performance}
In this section, we describe all experimentation we conducted regarding our domination score and the community detection algorithms. All experiments conducted on a Laptop with AMD Ryzen 7 PRO 5850U CPU@1.90 GHz with 32GB RAM and SSD Hard Drive.

We used the publicly available Aminer dataset\footnote{https://www.aminer.org/aminernetwork}~\cite{Tang:08KDD}, from which we used the co-authorship network that contains 1,712,433 authors and 4,258,615 collaboration relationships. Each author has the following attributes: Index ID, Name, Affiliations, Publication Count, Citation Number, H-Index, P-Index with equal A-index of this author, the P-index with unequal A-index of this author and Research Interests. We used the following four numerical attributes in our experiments: Publication Count (PC), Citation number (CN), H-Index (HI), P-Index (PI)~\cite{pnas.1220184110}

For our community search algorithm experiments, we randomly selected 100 nodes from the top 1\% of the entire dataset, based on each node's domination score. This selection resulted in nodes with domination scores ranging from 1,674,125 to 1,712,238

To evaluate the performance of our algorithm, we conducted experiments using these 100 randomly selected nodes.
We implemented two distinct search strategies: the hop-search approach and the random walk search approach.

For the hop-search approach, we utilized two specific parameters representing the distance from the initial node: 1-hop and 2-hop distances. This allowed us to examine the algorithm's performance at these two levels.

For the random walk search approach, we systematically varied two parameters: path length and the number of iterations. We tested all combinations of the following values for path length: 10, 20, 30, 40, 50, 60, and 70 steps, and for iterations: 10, 20, 30, 40, 50, 60, and 70 iterations. This comprehensive exploration enabled us to thoroughly assess the influence of these parameters on the algorithm's effectiveness.

The results from both search strategies were then compared on two primary metrics: runtime and community similarity. The runtime analysis focused on determining whether the random walk search approach was faster than the hop-search approach, and if so, quantifying the extent of this speed up. The similarity analysis involved evaluating how closely the communities identified by the random walk search approach matched those identified by the hop-search approach. This comparison provided insights into the effectiveness and efficiency of the random walk search in identifying community structures relative to the hop-search method.

The source code for all algorithms presented in this paper is available at \url{https://github.com/ngeorgiadis/community-search}.

\subsection{Community Similarity Analysis}

\begin{figure}[!hbt]
    \centering
    \includegraphics[width=0.7\textwidth]{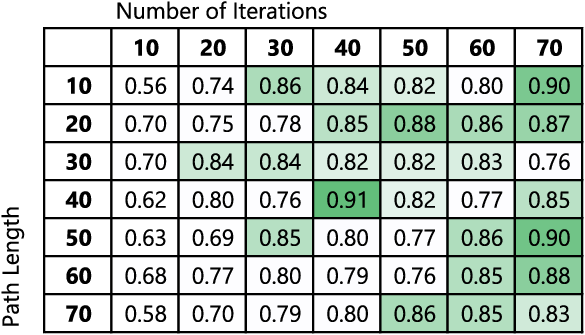}
    \caption{Average similarity of the communities produced after running variations of the random walk algorithm compared with the 2-hop search algorithm}\label{fig:avg-similarity}
\end{figure}

\begin{figure}[!hbt]
    \centering
    \includegraphics[width=0.8\textwidth]{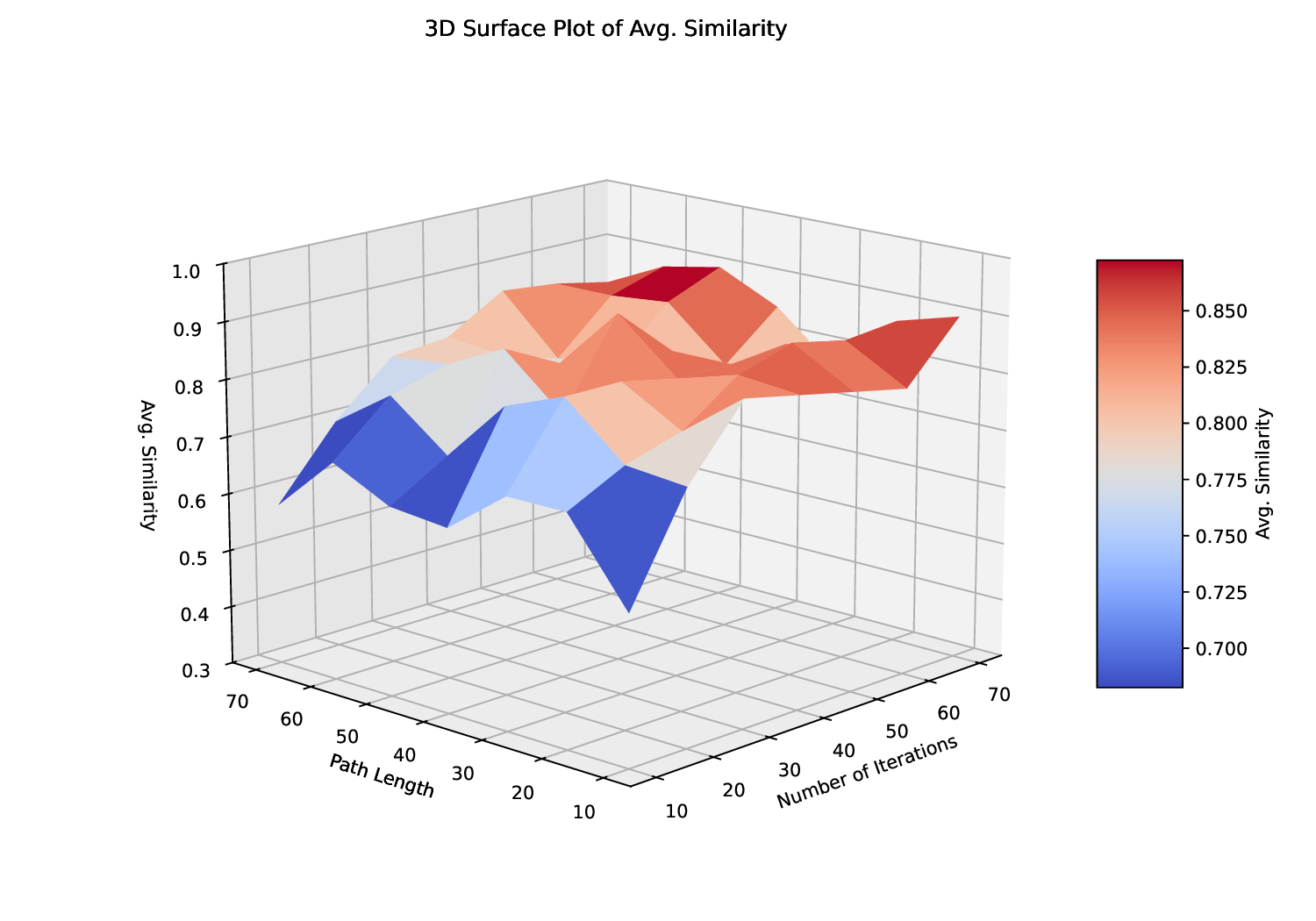}
    \caption{Surface plot showing the average similarity of the communities produced after running variations of the random walk algorithm}\label{fig:avg-similarity-3d}
\end{figure}

Figure~\ref{fig:avg-similarity} and Figure~\ref{fig:avg-similarity-3d} present the average similarity between communities identified by the 2-hop search algorithm and various configurations of the random walk search algorithm. The primary objective of this analysis is to evaluate how closely the communities detected by the random walk method approximate those found using the 2-hop approach, by varying the parameters of the random walk. Specifically, we explore the impact of the random walk path length (\textit{p}) and the number of iterations (\textit{w}).

The similarity metric used in this analysis quantifies the overlap between the node sets in the communities generated by the two methods. A higher similarity score indicates a greater resemblance between the communities identified by the random walk algorithm and those found by the 2-hop search. This allows us to assess the extent to which the random walk method can replicate the results of the 2-hop approach.

As shown in Figures~\ref{fig:avg-similarity},~\ref{fig:avg-similarity-3d}, certain combinations of path length and iterations yield high similarity scores suggesting that with appropriate tuning the random walk search can identify communities that are very similar to those discovered by the 2-hop method. These results demonstrate the flexibility and effectiveness of the random walk approach in detecting high-quality communities while offering potential improvements in computational efficiency. By identifying the optimal random walk configurations, we can achieve a balance between accuracy and runtime performance making this method a viable alternative for large-scale networks.

\subsection{Runtime Analysis of Random Walk Variations}
\begin{figure}[!hbt]
    \centering
    \includegraphics[width=0.7\textwidth]{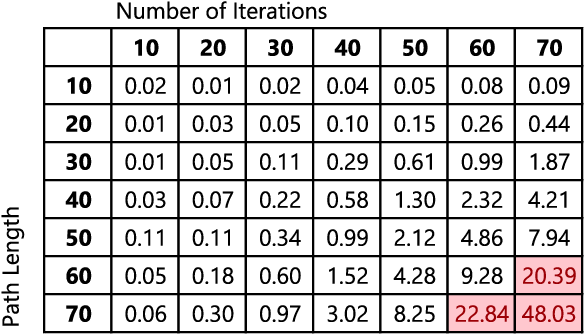}
    \caption{Mean runtime of the variations of the random walk algorithm compared with the 2-hop runtime in seconds }\label{fig:mean-runtime-vs-2-hop}
\end{figure}

\begin{figure}[!hbt]
    \centering
    \includegraphics[width=0.8\textwidth]{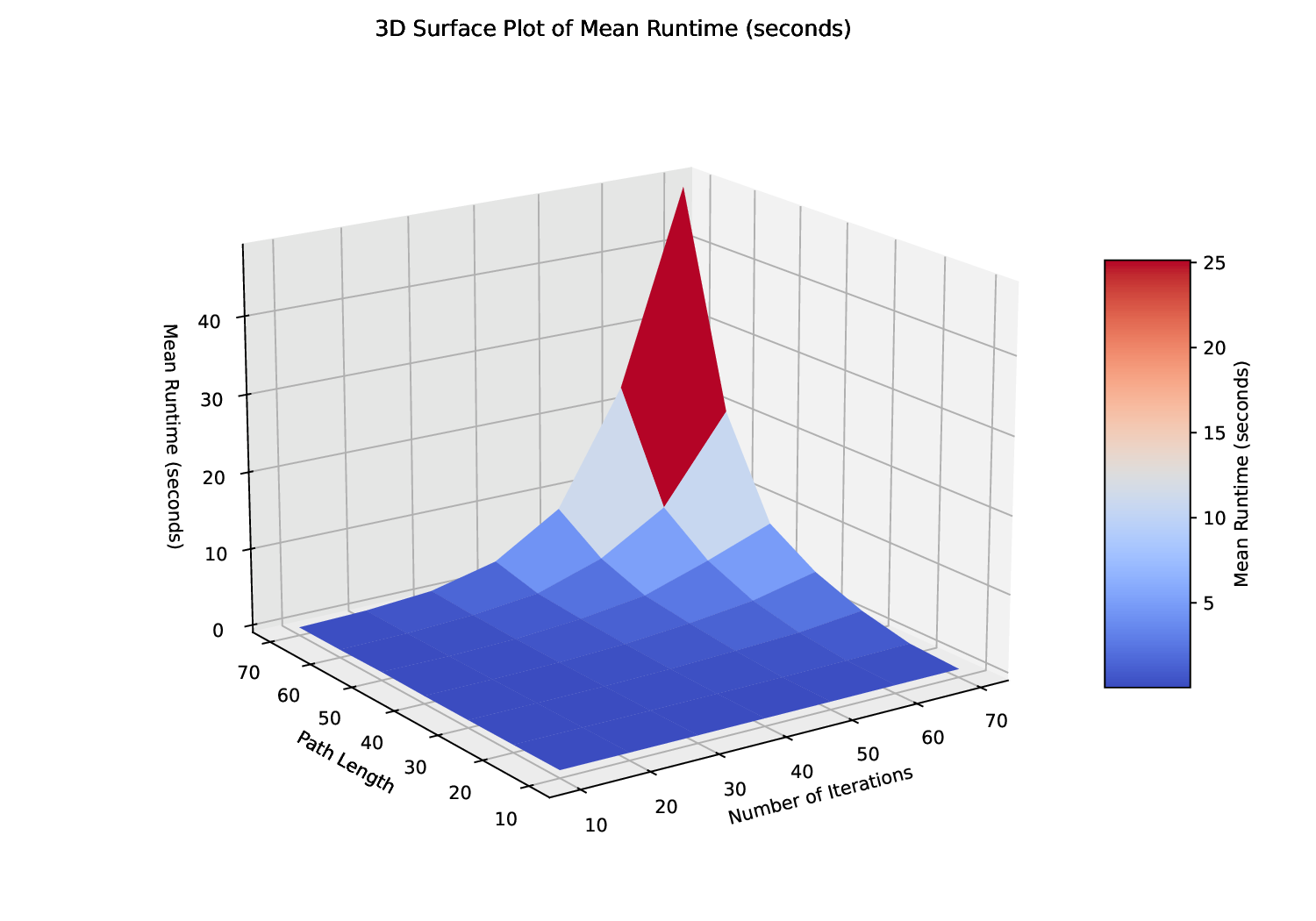}
    \caption{Surface plot showing the mean runtime of the variations of the random walk algorithm}\label{fig:mean-runtime-3d}
\end{figure}

Figure~\ref{fig:mean-runtime-vs-2-hop} and Figure~\ref{fig:mean-runtime-3d} illustrate the mean runtime of the random walk search algorithm for different configurations of path length (\textit{p}) and the number of iterations (\textit{w}). The purpose of those figures is to examine how changes in these parameters impact the algorithm's computational efficiency.

As expected increasing the path length or the number of iterations results in longer runtimes. This is because a larger path length requires the algorithm to explore a greater portion of the graph while more iterations increase the number of times the random walk is executed. The figure provides a clear visualization of the trade-off between parameter tuning and runtime performance.

This runtime analysis is crucial for selecting optimal random walk configurations that balance efficiency and accuracy. It allows us to understand the computational cost associated with deeper exploration of the graph controlled by the path length and with more extensive sampling determined by the number of iterations.

Additionally, we compare the mean runtime of the random walk search algorithm with the 2-hop search algorithm. This comparison is essential to demonstrate the computational efficiency of the random walk approach relative to the 2-hop method. Figures~\ref{fig:mean-runtime-vs-2-hop},~\ref{fig:mean-runtime-3d}, show the runtime for various random walk configurations in seconds. The runtime of the 2-hop search algorithm is \textbf{14.43 seconds} and serves as a baseline for comparison. As seen in the runtime table, certain configurations of the random walk search algorithm, particularly those with shorter path lengths and fewer iterations, exhibit significantly faster runtimes compared to the 2-hop search. For example, a path length of 30 and 30 iterations result in a runtime of \textbf{0.11 seconds}, which is orders of magnitude faster than the 2-hop method.

However, for configurations with longer path lengths and higher iterations, such as a path length of 70 with 70 iterations, the runtime increases to \textbf{48.03 seconds}, which is significantly higher than the 2-hop runtime. This emphasizes the importance of parameter tuning in random walk methods to strike a balance between computational efficiency and accuracy.

This analysis highlights the potential of the random walk search algorithm to provide faster results while maintaining competitive community detection quality, particularly in large-scale networks where computational resources are a constraint. The comparison with the 2-hop search runtime further solidifies the efficiency advantages of the random walk approach under optimal parameter settings.

\subsection{Community Density Analysis}

\begin{figure}[!hbt]
    \centering
    \includegraphics[width=0.7\textwidth]{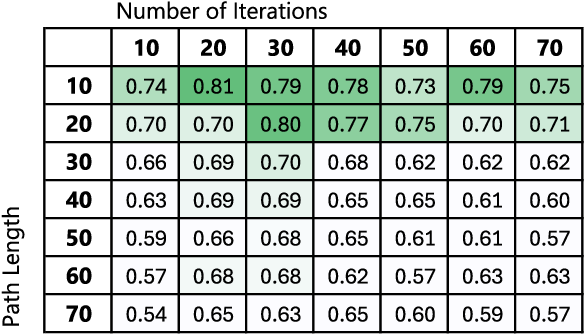}
    \caption{Mean density of the top-1 community}\label{fig:mean-top1-density}
\end{figure}

\begin{figure}[!hbt]
    \centering
    \includegraphics[width=0.8\textwidth]{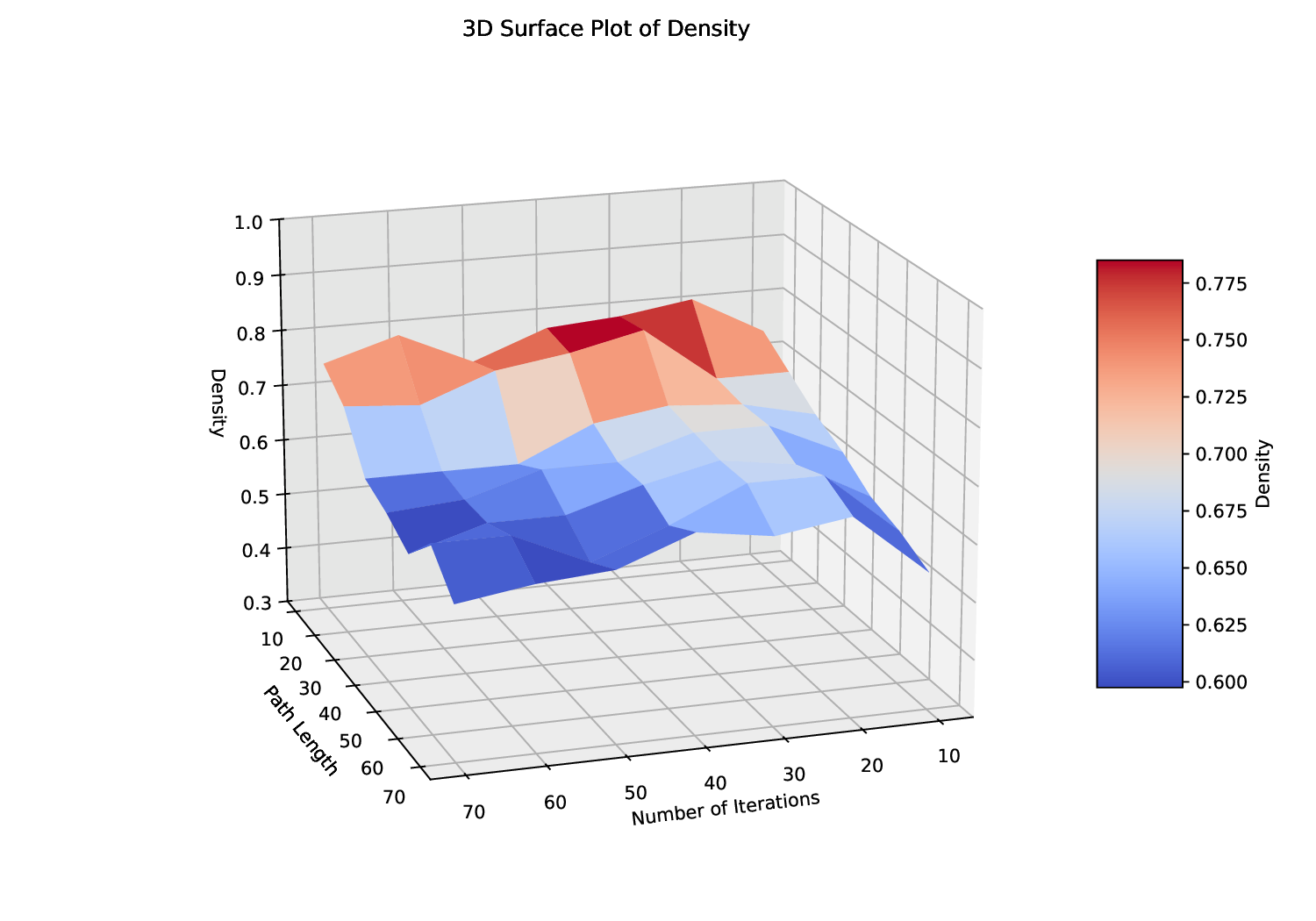}
    \caption{Surface plot showing the mean density of the top-1 community}\label{fig:mean-top1-density-3d}
\end{figure}

Figure~\ref{fig:mean-top1-density} and Figure~\ref{fig:mean-top1-density-3d} show the mean density of the top-ranked community detected by the random walk search algorithm. Density is a measure of how well-connected the nodes in a community are, where higher density indicates a more cohesive community structure. We use the following definition of density, which is the ratio of the number of actual edges in the community to the number of possible edges in that community. The mean density for the 1-hop and 2-hop search methods was \textbf{0.84} and \textbf{0.65} respectively. The table demonstrates that the random walk algorithm can identify communities with comparable or higher density than the 2-hop search, particularly for certain parameter configurations. The highest mean density of \textbf{0.81} and \textbf{0.80} was achieved with a path length of 10 and 20 iterations, and a path length of 20 and 30 iterations, respectively. These results indicate that the random walk algorithm can produce communities with strong internal cohesion, similar to those identified by the 2-hop search method. This analysis reinforces the robustness of the random walk method, demonstrating that it can produce high-quality communities with strong internal cohesion, while also offering faster runtimes as shown in the previous figures. This makes the random walk search algorithm not only efficient but also effective in community detection tasks.

\subsection{$\beta$-Index Analysis}
\begin{figure}[!hbt]
    \centering
    \includegraphics[width=0.7\textwidth]{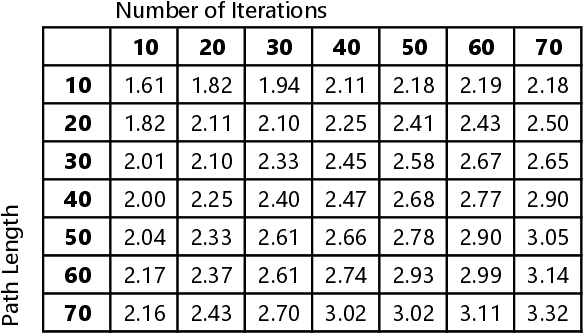}
    \caption{$\beta$-index of the top-1 community}\label{fig:mean-top1-b-index}
\end{figure}

\begin{figure}[!hbt]
    \centering
    \includegraphics[width=0.8\textwidth]{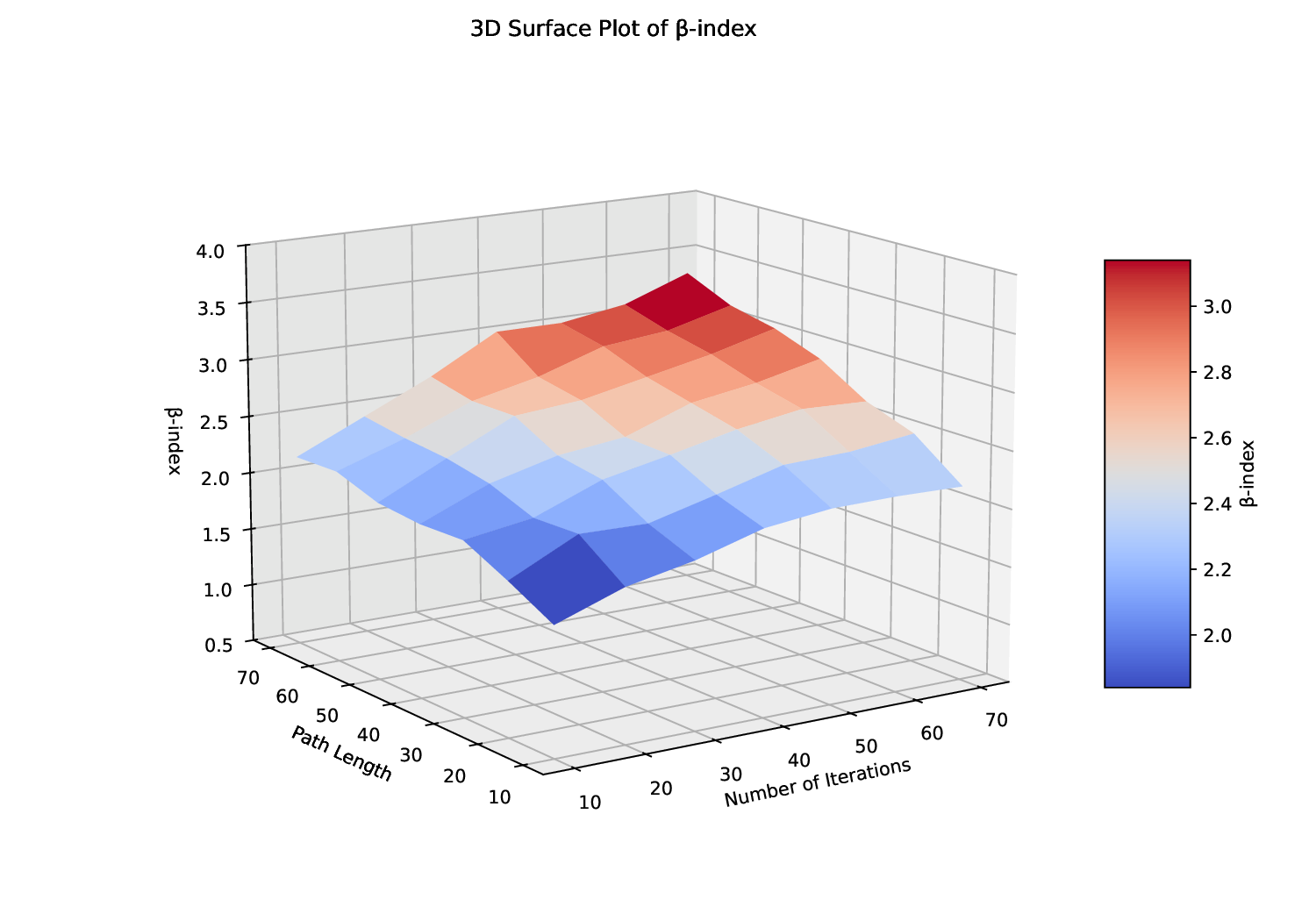}
    \caption{Surface plot showing the $\beta$-index of the top-1 community}\label{fig:mean-top1-b-index-3d}
\end{figure}

The $\beta$-index~\cite{Rodrigue2024} is similar to the density metric but provides a different view of community structure by considering both the number of edges and the number of nodes in the community. It is defined as the ratio of the number of edges to the number of nodes in the community. A higher $\beta$-index indicates a more connected community, where nodes are well-connected relative to their size. An index smaller than 1 indicates that the community is sparse, while an index greater than 1 indicates a dense community. If the index is equal to 1 then the community contains one circle. Figures~\ref{fig:mean-top1-b-index} and~\ref{fig:mean-top1-b-index-3d} show the mean $\beta$-index of the top-ranked community detected by the random walk search algorithm. The mean $\beta$-index for the 1-hop and 2-hop search methods was \textbf{2.65} and \textbf{3.53} respectively. We observe that the random walk algorithm reaches the level of the 1-hop search when the path length is 50 or greater and the number of iterations is 30 or greater. At the highest experimental parameter settings, the random walk algorithm achieves a mean $\beta$-index of \textbf{3.32}, which is comparable but not better than the 2-hop search method.

\subsection{Comparison of Community Structures: 2-Hop Search vs. Random Walk}
The comparison between the 2-hop search and random walk algorithms reveals a trade-off between computational efficiency and community detection quality. The random walk method with appropriate parameter tuning can closely approximate the community structures identified by the 2-hop search algorithm as shown by high similarity scores. Moreover, the random walk algorithm often detects communities with comparable or better density, demonstrating its ability to uncover cohesive and well-connected subgraphs. These results highlight the random walk's adaptability, making it a competitive alternative for applications requiring high-quality community detection.

Regarding efficiency, the random walk algorithm significantly outperforms the 2-hop method in runtime under moderate configurations, such as shorter path lengths and fewer iterations. This makes it particularly advantageous for large-scale networks where computational resources are limited. However, extensive parameter settings (longer path lengths or more iterations) can result in runtimes exceeding those of the 2-hop approach. Thus, the random walk algorithm provides a flexible solution enabling users to balance computational cost and detection quality based on their specific requirements.

\subsection{Random Walk Algorithm Parameter Tuning}
The optimal parameters for the random walk algorithm depend on balancing runtime efficiency and community detection quality. For most applications, a moderate path length of 30--50 steps and 20--40 iterations provides a strong balance. These settings allow the algorithm to explore the network effectively while maintaining manageable computational costs. Shorter path lengths and fewer iterations may sacrifice some accuracy but significantly improve runtime, making them suitable for very large networks or time-sensitive tasks. In contrast, longer paths (60--70 steps) or higher iterations (50--70) improve robustness and quality but come with a steep increase in runtime, which should be reserved for scenarios where computational resources are less constrained.

A recommended configuration for balanced performance is a path length of 40 and 30 iterations. This combination achieves near-optimal similarity to the 2-hop method while being significantly faster. For faster execution with acceptable community quality, a path length of 30 and 20 iterations is a good choice. Adapting parameters dynamically based on network properties, such as size or density, can further optimize performance, ensuring the algorithm remains effective and efficient across diverse datasets.

\subsection{Evaluation Metrics}
We use the domination score to evaluate and rank the communities created by our algorithms. In the first step of each algorithm, we generate a subgraph derived from the whole graph to search for communities. Within this subgraph, we find the node with the highest domination score, which we call \( MAX_{dom} \). This highest score serves as a reference for our ranking metric. The idea is to measure how close each community's nodes are to the \( MAX_{dom} \) in terms of their domination scores. Since the domination score is based on the node's attributes, communities with scores closer to \( MAX_{dom} \) are considered better. This ensures that we prioritize communities with attributes similar to the top-scoring node in the initial subgraph. The metric is expressed in Equation~\ref{eq:community-deviation}.

\begin{equation}
    \centering
    \sigma = \sqrt{\frac{\sum_{i=1}^{N} |dom_i-MAX_{dom}|^2}{N}}
    \label{eq:community-deviation}
\end{equation}

\subsection{Communities Examples}

\begin{figure}[!hbt]
    \centering
    \includegraphics[width=0.7\textwidth]{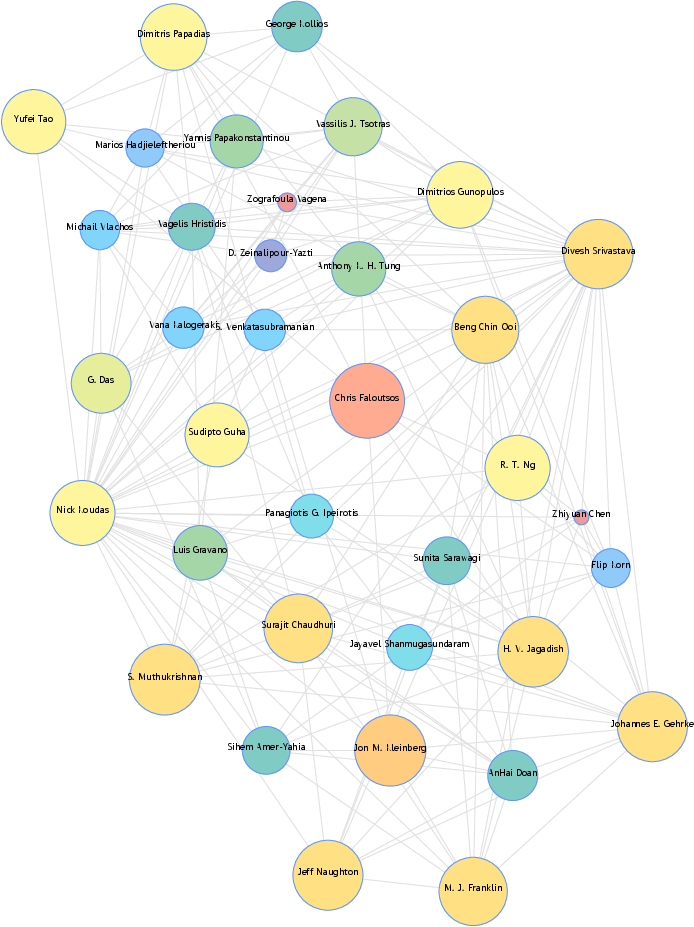}
    \caption{Example of 1-hop community}\label{fig:example_hop1}
\end{figure}

\begin{figure}[!hbt]
    \centering
    \includegraphics[width=0.7\textwidth]{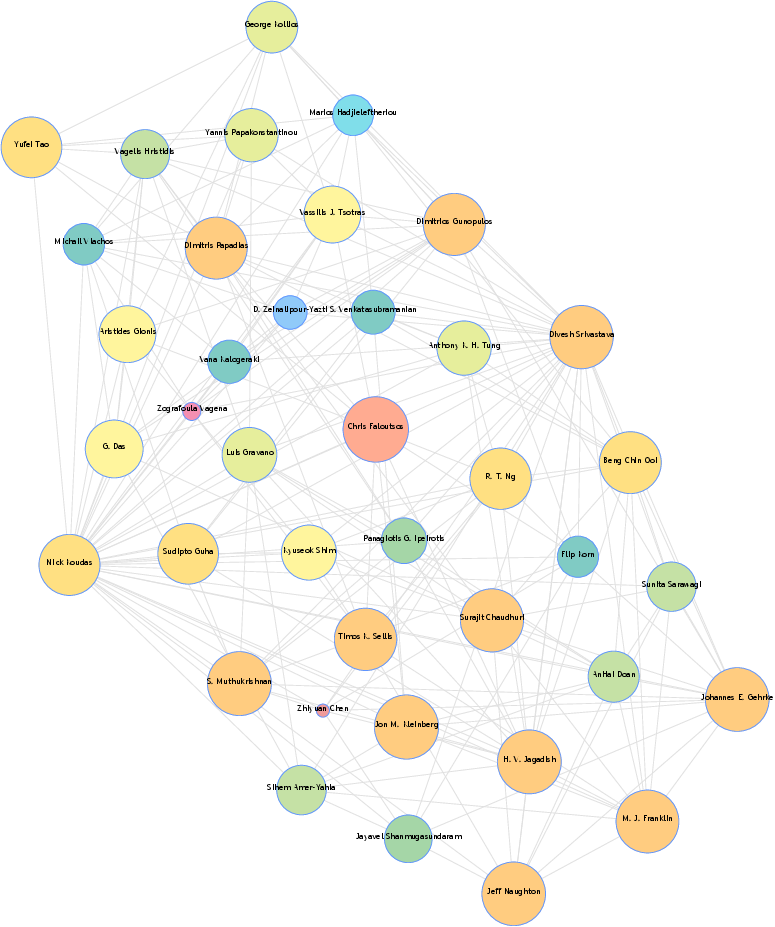}
    \caption{Example of 2-hop community}\label{fig:example_hop2}
\end{figure}

\begin{figure}[!hbt]
    \centering
    \includegraphics[width=0.7\textwidth]{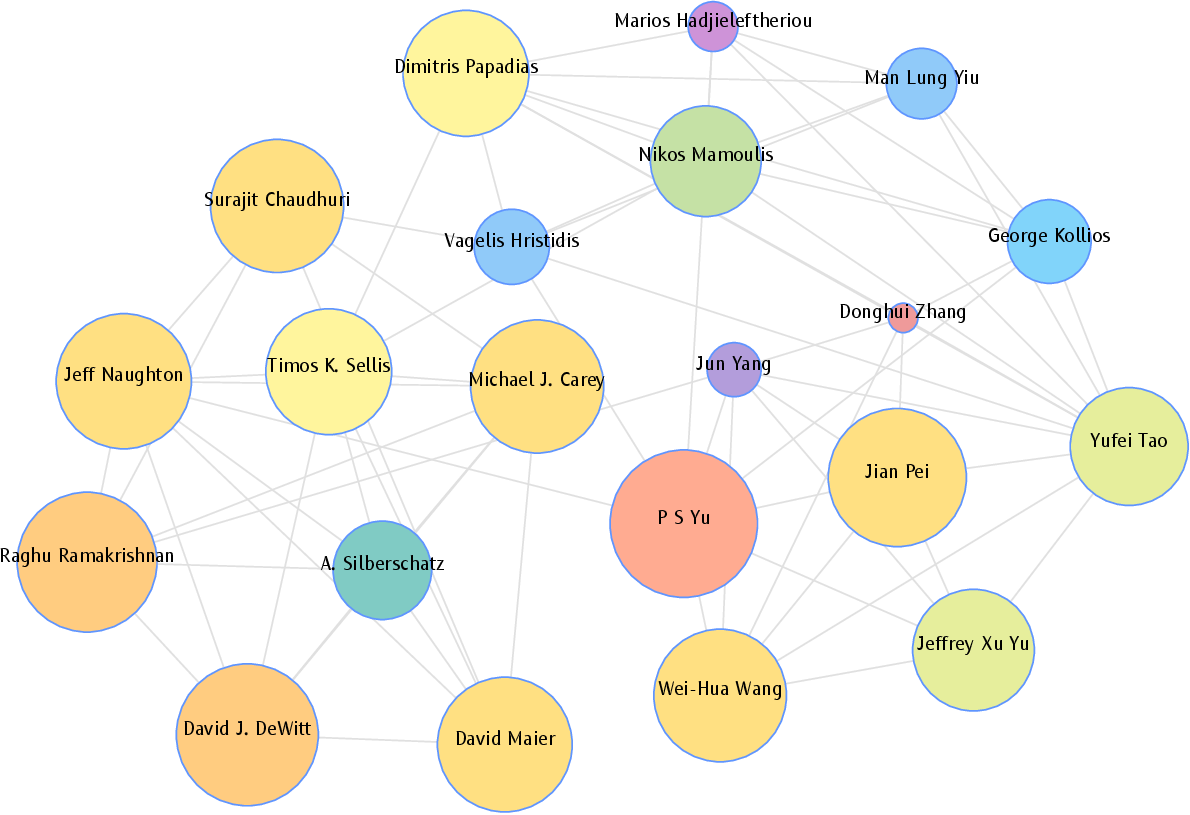}
    \caption{Example of Random Walk community}\label{fig:example_rw}
\end{figure}

Figures~\ref{fig:example_hop1},~\ref{fig:example_hop2} and~\ref{fig:example_rw} show the communities produced by the hop search and random walk search for a specific node (Node ID:456112). We can see that the communities produced by the hop search and random walk search are similar in terms of structure and node composition. The random walk search algorithm is able to identify a community that closely resembles the one found by the hop search method, demonstrating its effectiveness in community detection tasks. The visualizations also highlight the cohesive nature of the communities, with nodes being well-connected within their respective groups. This further supports the idea that the random walk search can be a viable alternative to traditional methods like hop search for community detection in large-scale networks.
%
%

%
%
\section{Conclusion}\label{sec:conclusions}
In this paper, we introduced a novel approach for community search in attributed networks, considering hop-based and random walk-based algorithms. Our techniques balance the structural connectivity of the network with node attribute similarities to identify cohesive and meaningful communities. By calculating a domination score and applying k-core extraction, we ensure that the communities discovered are both well-connected and attribute-rich.

Through comprehensive experiments, we demonstrated the effectiveness of our approach using real-world datasets. The random walk-based method was shown to outperform the traditional 2-hop search in terms of runtime, particularly for shorter path lengths and fewer iterations, while maintaining comparable or better community quality, as indicated by community similarity and density metrics. This makes the random walk-based method a compelling alternative for large-scale networks where computational efficiency is paramount.

Our analysis of runtime and community density confirmed that the proposed method is not only efficient but also capable of detecting highly cohesive communities. Additionally, by tuning the parameters of the random walk (path length and number of iterations), the algorithm can achieve a favorable trade-off between accuracy and performance.

There are several potential directions for future research. First, the scalability of the algorithm could be further improved by exploring more advanced approximation techniques or distributed computing approaches. Additionally, the algorithm could be adapted to dynamic networks where communities evolve over time. Finally, extending the approach to multilayer networks, where nodes and edges may span across different dimensions or domains, represents another promising avenue for enhancing the versatility of the proposed method.

%
%
\bibliographystyle{plain}
\bibliography{bibliography}

\end{document}